\documentclass[11pt]{article}
\pdfoutput=1
\usepackage{amsmath, amsfonts, amssymb}
\usepackage{comment}
\usepackage{graphicx}
\usepackage{psfrag}
\usepackage{amsthm}
\usepackage[usenames,dvipsnames,svgnames,table]{xcolor}
\usepackage{enumerate}
\usepackage{tcolorbox}
\usepackage{mathtools}

\usepackage{soul}
\usepackage{subfig}
\usepackage{mathrsfs}  
\usepackage{a4wide}
\usepackage{booktabs}
\usepackage{tikz}
\usepackage{tikz-cd}
\usepackage{makecell}

\usepackage{color}
\definecolor{dark-gray}{gray}{0.20}
\definecolor{gray}{gray}{0.30}
\definecolor{light-gray}{gray}{0.80}
\definecolor{dark-red}{rgb}{0.7,0,0}
\definecolor{dark-green}{rgb}{0.1,0.4,0}
\definecolor{dark-blue}{rgb}{0.3,0.3,0.7}
\definecolor{light-blue}{rgb}{0.8,0.8,1}
\definecolor{swamp}{RGB}{240, 199, 197}

\usepackage{pifont}
\usepackage{setspace}

\newcommand{\be}{\begin{equation}}
\newcommand{\ee}{\end{equation}}

\def\be{\begin{equation}}
\def\ee{\end{equation}}
\def\bea{\begin{eqnarray}}
\def\eea{\end{eqnarray}}

\newcommand{\dd}{\mathrm{d}}

\def\simleq{\; \raise0.3ex\hbox{$<$\kern-0.75em
		\raise-1.1ex\hbox{$\sim$}}\; }
\def\simgeq{\; \raise0.3ex\hbox{$>$\kern-0.75em
		\raise-1.1ex\hbox{$\sim$}}\; }

\numberwithin{equation}{section}

\usepackage{jheppub} 
\hypersetup{
	colorlinks=true,
	linkcolor=dark-blue,
	citecolor=dark-red,
	urlcolor=dark-green,
	linktoc=page,
	pageanchor=false
}

\title{
\centering The LVS Parametric Tadpole Constraint}

\author{Xin Gao$^1$,}
\author{Arthur Hebecker$^2$,}
\author{Simon Schreyer$^2$,}
\author{ and Gerben Venken$^2$}

\affiliation{$^1$ College of Physics, Sichuan University, Chengdu, 610065, China}
\affiliation{$^2$ Institute for Theoretical Physics, Heidelberg University,\\
	Philosophenweg 19, 69120 Heidelberg, Germany} 
\emailAdd{xingao@scu.edu.cn}
\emailAdd{a.hebecker@thphys.uni-heidelberg.de}
\emailAdd{s.schreyer@thphys.uni-heidelberg.de}
\emailAdd{g.venken@thphys.uni-heidelberg.de}

\abstract{The large volume scenario (LVS) for de Sitter compactifications of the type IIB string is, at least in principle, well protected from various unknown corrections. The reason is that, by construction, the Calabi-Yau volume is exponentially large. However, as has recently been emphasised, in practice the most explicit models are rather on the border of parametric control. We identify and quantify parametrically what we believe to be the main issue behind this difficulty. Namely, a large volume implies a shallow AdS minimum and hence a small uplift. The latter, if it relies on an anti-D3 in a throat, requires a large negative tadpole. As our main result, we provide a simple and explicit formula for what this tadpole has to be in order to control the most dangerous corrections. The fundamental ingredients are parameters specifying the desired quality of control.
We comment on the interplay between our constraint and the tadpole conjecture. We also discuss directions for future work which could lead to LVS constructions satisfying the tadpole constraint with better control, as well as further challenges that may exist for the LVS. Our formula then represents a very concrete challenge for future searches for and the understanding of relevant geometries.
}

\begin{document}

\makeatletter
\let\old@fpheader\@fpheader

\makeatother

\maketitle


\section{Introduction}\label{intro}
The construction of metastable de Sitter vacua in string theory is challenging and, in spite of many years of work, the focus has remained on two close cousins: the KKLT \cite{Kachru:2003aw} and the LVS \cite{Balasubramanian:2005zx} proposals. Recently, fundamental doubts about the very existence of de Sitter space in quantum gravity (see e.g.~\cite{Danielsson:2018ztv, Obied:2018sgi, Garg:2018reu, Ooguri:2018wrx}) have triggered intense scrutiny of these models. KKLT has arguably survived attacks related to the stability question of the anti-D3 uplift \cite{Bena:2009xk,Bena:2011wh,Michel:2014lva,Blaback:2019ucp} and the gaugino condensate \cite{Moritz:2017xto,Carta:2019rhx,Gautason:2019jwq,Hamada:2019ack,Kachru:2019dvo,Hamada:2021ryq}. It is too early to judge how it will fare in view of the singular-bulk problem \cite{Gao:2020xqh}, which arises since one is forced to glue a large throat into a fairly small Calabi-Yau \cite{Carta:2019rhx}. In view of this uncertainty (see \cite{Carta:2021lqg,Demirtas:2021nlu} for some the recent discussion), it is maybe natural to shift the focus on the LVS, where the singular-bulk problem does, at first sight, not arise \cite{Gao:2020xqh} because the volume can be made exponentially large.

However, as recently emphasised in \cite{Junghans:2022exo}, the problem does not entirely disappear either. The reason is that at large volume the AdS minimum of the LVS becomes shallow, requiring a small uplift and hence a strongly warped throat. The latter comes with a large positive D3 charge and hence requires a large negative D3 charge of the embedding Calabi-Yau orientifold. This requirement has been emphasised as a potential problem in \cite{Bena:2020xrh} and argued to be a solvable issue in \cite{Crino:2020qwk}, where the focus was on a concrete model. Most recently, this model has in turn been challenged in \cite{Junghans:2022exo} on the basis that all the different corrections can not be made sufficiently small at the same time.

Our purpose in this short note is to derive a simple formula quantifying what we believe to be the most serious obstacle to defending the LVS against such skepticism. We call our result the `LVS parametric tadpole constraint', as it quantifies the amount of negative tadpole which the geometry must possess to be in perturbative control. The desired quality of control is an input to this formula.

Our result will follow from two different corrections to the LVS potential, both with the same parametric behaviour.
One is related to warping and hence similar in spirit to the singular-bulk problem, the other to higher $F$-terms \cite{deAlwis:2005tg,Cicoli:2013swa,Ciupke:2015msa}. As we will see, this leads to a simple, strong and conceptually clear bound on controlled LVS models with de Sitter uplift, deriving from the requirement to keep warping and higher $F$-term effects small at the same time. By comparison, the recent analysis of \cite{Junghans:2022exo} used these conditions together with a number of conditions related to further corrections (see e.g.~\cite{Conlon:2010ji,Antoniadis:2019rkh,Conlon:2009kt,Weissenbacher:2019mef,Grimm:2013bha,Grimm:2015mua,Junghans:2014zla,vonGersdorff:2005bf,Berg:2005ja,Berg:2007wt,Cicoli:2007xp,Cicoli:2008va,Burgess:2020qsc,Cicoli:2021rub,Minasian:2015bxa}) to collectively constrain the LVS. These analytical constraints were then applied to a specific Calabi-Yau orientifold.
The most important of those further corrections are probably certain logarithmic contributions to the Kahler potential, which are not yet unambiguously established. We will comment on this in slightly more detail below. Many of the remaining corrections are interesting but probably not as dangerous. We will discuss them only briefly in what follows.

We note that in \cite{Gao:2022uop} we provide a general discussion of loop corrections to the Kahler moduli Kahler potential in type-IIB compactifications \cite{vonGersdorff:2005bf,Berg:2005ja,Berg:2007wt,Cicoli:2007xp,Cicoli:2008va}. We will in particular discuss at which order in $\alpha'$, $g_s$, and cycle volumes loop corrections can appear and what their potential impact on proposed compactifications, such as the LVS, is. 

The rest of the present paper is organized as follows: We list the most important LVS relations in Sect.~\ref{sum}, we derive the proposed key equation for the required minimal tadpole in Sect.~\ref{cons}. Then, in Sect.~\ref{sec:tadpoleconjecture}, we discuss how one may try to strengthen the LVS proposal in the future in spite of the presented constraint.  We also discuss further challenges and comment on the interplay of our constraint with the tadpole conjecture \cite{Bena:2020xrh, Bena:2021wyr, Plauschinn:2021hkp, Lust:2021xds, Bena:2021qty}.

\section{Summary of basic equations}\label{sum}

To set the scene, we recall the LVS set-up. One compactifies IIB string theory to 4D on a Calabi-Yau orientifold $X$ with flux and warping. The geometry has at least a big 4-cycle $\tau_b$ and a small 4-cycle $\tau_s$. The corresponding Kahler potential reads \cite{Balasubramanian:2005zx,Becker:2002nn}
\begin{equation}
    K=-2\ln\left(\mathcal{V} + \frac{\xi}{2 g_s^{3/2}}\right) = -2 \ln \left( \tau_b^{3/2}-\kappa_s\tau_s^{3/2}-\frac{\chi\,\zeta(3)}{4(2\pi)^3g_s^{3/2}} \right)\,.
\end{equation}
Here ${\cal V} = \tau_b^{3/2}-\kappa_s \tau_s^{3/2}$ is the Calabi-Yau volume in units of $l_s=2\pi\sqrt{\alpha'}$ and after a rescaling to the 10d Einstein frame. The constant $\kappa_s$ is related to the triple intersection number of $\tau_s$ and an analogous constant in front of $\tau_b$ has been absorbed by rescaling $\tau_b$.  The string coupling is $g_s$ and $\chi$ is the Euler number of the Calabi-Yau.
The superpotential is given by
\begin{equation}
    W = W_0 + A_s \text{e}^{-a_s T_s}\,,
\end{equation}
with $W_0$ due to flux and the nonperturbative correction coming from either ED3-branes ($a_s=2 \pi$) or gaugino condensation on D7-branes ($a_s$ depends on the gauge group, e.g.~$a_s=\pi/3$ for $SO(8)$). The prefactor $A_s$ is model-dependent. We have absorbed the complex structure Kahler potential in a multiplicative rescaling of $W_0$ and $A_s$.

This yields the pure LVS scalar potential (in 4d Planck units)
\begin{equation}
    V = \frac{4a_s^2|A_s|^2g_s\sqrt{\tau_s}\text{e}^{-2a_s\tau_s}}{3\kappa_s\mathcal{V}} - \frac{2a_s|A_s| g_s \tau_s |W_0| \text{e}^{-a_s \tau_s}}{\mathcal{V}^2} + \frac{3\xi |W_0|^2}{8\sqrt{g_s}\mathcal{V}^3}\,,
    \label{pureLVSpotential}
\end{equation}
which is minimized by
\be
    \label{vts}
    \mathcal{V}=\frac{3\kappa_s|W_0|\sqrt{\tau_s}}{4a_s |A_s|}\text{e}^{a_s\tau_s}\,\,\,,\qquad
    \qquad
    \tau_s = \frac{\xi^{2/3}}{(2\kappa_s)^{2/3}g_s} +\mathcal{O}(1)\,,
\ee
leading to an AdS vacuum at
\begin{equation}
    V_\text{AdS} = - \frac{3\kappa_s g_s \sqrt{\tau_s} |W_0|^2}{8a_s\mathcal{V}^3}\,.
    \label{vads}
\end{equation}
As noted in the Appendix of \cite{Hebecker:2012aw} this value is parametrically suppressed with respect to the individual terms in $V$ by a factor of $1/\tau_s$ or $g_s$. This has been emphasised and studied in detail in \cite{Junghans:2022exo} under the name `non-perturbative no-scale structure'.

One now assumes that a Klebanov-Strassler throat is present and places an anti-D3 brane at its tip. This adds a metastable uplift \cite{Kachru:2003aw}
\begin{equation}
    V_{\text{uplift}} = V_{\overline{D3}} = 2 T_{D3} e^{4 A(0)}
\end{equation}
to the scalar potential. Here $T_{D3}$ is the D3-brane tension in 4D Einstein frame and $e^{A(0)}$ is the warp factor at the tip of the throat. 
To determine this warp factor, we neglect the difficult issue of backreaction \cite{Bena:2018fqc,Blumenhagen:2019qcg} since it is expected to make the uplift smaller. Hence we are being conservative by disregarding it. We rather use the standard Klebanov-Strassler geometry, glued into the conical region of a compact Calabi-Yau. We are careful to keep all ${\cal O}(1)$ factors except for those which are unknown since they are related to the specific Calabi-Yau (cf.~App.~\ref{upliftpotential} for details). Including in particular the corrected volume scaling of~\cite{Kachru:2003sx}, we find
\begin{equation}
    e^{4 A(0)} = \left(\frac{g_s^{3/2} \mathcal{V}}{\mathcal{V}_0 }\right)^{2/3} \frac{3\, 2^{2/3}\pi}{a_0} \, \frac{N}{g_sM^2} e^{-\frac{8\pi K}{3g_s M}}\,.
\end{equation}
Here $a_0\approx 0.71805$ is a numerical constant related to the explicit Klebanov-Strassler geometry \cite{Klebanov:2000hb}, $\mathcal{V}_0$ is the volume of the $T^{1,1}$ where the conifold is glued into the Calabi-Yau, and $M$ is the $F_3$ and $K$ the $H_3$ flux on the throat 3-cycles. The contribution of the flux in the throat to the D3-tadpole is $N = K M$. The final form of the uplift potential is then 
\begin{equation}
    V_\text{uplift} = \frac{\left( 3^2\,\pi^3\, 2^{22/3} \right)^{1/5}}{a_0} \frac{\text{e}^{-\frac{8\pi K}{3g_s M}}}{g_s M^2\mathcal{V}^{4/3}}\,.
    \label{vuplift}
\end{equation}
Jumping slightly ahead we note that the weaker volume suppression of the uplift as compared to the depth of the AdS potential in \eqref{vads} does not represent a problem. The matching of the two effects will be ensured by the non-trivial volume scaling of $|W_0|^2$ and of the exponential warping suppression in \eqref{vuplift}.

Before moving on to the derivation of the tadpole constraint, we have to discuss one of the so-called higher $F$-term corrections. It is important since it bounds the size of $W_0$ in terms of the Calabi-Yau volume\footnote{The resulting constraint is slightly stronger than the gravitino constraint $m_{3/2}\ll m_{KK}$ needed for a valid low-energy supergravity description \cite{Conlon:2005ki}. We thank Daniel Junghans for making us aware of this.}.
Concretely, this correction comes from eight derivative terms involving the field strength $G_3$ and takes the form \cite{Ciupke:2015msa,Junghans:2022exo}
\begin{equation}
    \delta V_F \sim \frac{W_0^4 g_s^{1/2}}{\mathcal{V}^{11/3}}\,.
    \label{higherfterms}
\end{equation}
Its ratio to the value of the AdS potential at the minimum \eqref{vads} is required to be small. To quantify this, we multiply this ratio by our control parameter $c_{W_0}$ and set the result equal to unity:
\begin{equation}
    1 = c_{W_0} \frac{16 a_s }{3 (2\kappa_s)^{2/3}\xi^{1/3}}\, \frac{W_0^2}{\mathcal{V}^{2/3}}\,.
    \label{cw0}
\end{equation}
For $c_{W_0}\gg 1$, higher terms in the superspace derivative expansion are suppressed \cite{deAlwis:2005tg, Cicoli:2013swa, Ciupke:2015msa}.

Let us briefly comment on the actual prefactor in \eqref{higherfterms}. In \cite{Ciupke:2015msa} the correction is determined to be $V_F=\hat{\lambda}W_0^4g_s^{1/2}\,\Pi_i\,t^i/\mathcal{V}^4$ where $t^i$ is some 2-cycle volume, $|\hat{\lambda}|$ can roughly be estimated to be of order $|\xi/\chi|\sim 10^{-3}$, and $\Pi_i$ are topological numbers estimated to be of $\mathcal{O}(10-100)$. The topological numbers can be computed in explicit examples \cite{Ciupke:2015msa,Cicoli:2016xae,Cicoli:2017axo}.\footnote{We thank Pramod Shukla for discussions regarding this point.} Note that in \eqref{higherfterms} we focused on the contribution of the largest 2-cycle $t_\ast\sim\mathcal{V}^{1/3}$.

\section{The LVS Parametric Tadpole Constraint}\label{cons}
The LVS tadpole constraint is derived in two steps: First, we require the uplift potential \eqref{vuplift} to be of the size of the potential in the AdS minimum \eqref{vads}.\footnote{
The relative fine tuning of the two terms can always be realised by adjusting $|W_0|$, which should not be problematic as long as $c_{W_0}$ has not been chosen too close to unity.
} 
This leads to
\begin{equation}
    \mathcal{V} = \frac{a_0}{(3^2 \pi^3 2^{22/3})^{1/5}}\left( \frac{3(2\kappa_s)^{2/3}\xi^{1/3}}{16 a_s} \right)^2 \frac{ g_s^{1/2}}{ c_{W_0} } g_sM^2 \, \text{e}^{\frac{8\pi}{3g_sM^2} N } \,,
    \label{tauV}
\end{equation}
where we replaced $|W_0|$ by \eqref{cw0}. Setting $\mathcal{V}$ in \eqref{vts} equal to \eqref{tauV} gives an implicit equation for $\tau_s=\tau_s(N)$ which can be solved to leading order by comparing the exponentials:\footnote{Note that replacing $W_0$ by the higher F-term constraint changes the exponential factor in \eqref{vts} to $\text{e}^{3a_s\tau_s/2}$.}
\begin{equation}
    a_s\tau_s =  \frac{16\pi N }{9 g_s M^2} + \ln(\mathcal{O}(1)) \,.
    \label{soltauN}
\end{equation}

In the second step, we bound the volume from below by demanding that we have a well-controlled solution. One argument is to require that the singular-bulk problem is avoided. This imposes \cite{Gao:2020xqh,Carta:2019rhx}
\begin{equation}
    1 \gg \frac{N}{\mathcal{V}^{2/3}}\,.
    \label{cN}
\end{equation}
However, as recently argued in \cite{Junghans:2022exo}, a related and more quantitative bound can be given by studying 10d  higher-curvature terms. Indeed, before strong warping leads to the formation of singular regions, it drives curvature corrections large. Specifically, consider a correction to the 4d Einstein-Hilbert term, analogous to BBHL \cite{Becker:2002nn}, arising from the interplay between the 10d $R_{10}^4$ term and a varying warp factor \cite{Junghans:2022exo}: 

One uses the warped metric ansatz 
\begin{equation}
    \dd s^2=\text{e}^{2A(y)}g_{\mu\nu}\dd x^\mu\dd x^\nu + \text{e}^{-2A(y)} \tilde{g}_{mn}\dd y^m\dd y^n
\end{equation}
together with the term
 \begin{equation}
     \frac{1}{g_s^{3/2}}\int\dd^{10}x \sqrt{-G}  \varepsilon^{ABM_1\dots M_8} \varepsilon_{ABN_1\dots N_8} R^{N_1N_2}_{~~~~~M_1M_2}\cdots R^{N_7N_8}_{~~~~~M_7M_8}
     \label{R4}
 \end{equation}
from the $R^4_{10}$ correction to the type-IIB action \cite{Becker:2002nn,Antoniadis:1997eg} in 10d Einstein frame. To derive the contribution to the 4d Einstein-Hilbert term, we focus on fluctuations of the type $g_{\mu\nu}=g_{\mu\nu}(x)$. We also disregard terms with derivatives of $A(y)$. This turns \eqref{R4} into
 \begin{equation}
     \frac{1}{g_s^{3/2}}\int_{\mathcal{M}_{10}}e^{2A(y)}R\wedge R\wedge R\wedge R\wedge e \wedge e\,,
 \end{equation}
where $R$ is the Riemann tensor of the unwarped Calabi-Yau compactification with metric $(g_{\mu\nu},\tilde{g}_{mn})$ and $e$ denotes the vielbein 1-form. Since we are now working with a product geometry, the contribution to the 4d Einstein-Hilbert term can be made explicit as
\begin{align}
     &\frac{1}{g_s^{3/2}}\int\dd^4x R_4 \int_X R_6\wedge R_6\wedge R_6\, \Big(1+(\text{e}^{2A(y)}-1)\Big) \nonumber
     \\ \approx &\frac{1}{g_s^{3/2}}\int d^4 x R_4 \left(\chi(X) + \frac{\chi(X) N}{\mathcal{V}^{2/3}}
    \right).
\end{align}
Here we assumed that the bulk of the Calabi-Yau is at weak warping, $e^{2A}\simeq 1$. We may then use the fact that, in the 10d string frame, $(e^{-4A}-1)$ is the solution of a Poisson equation with the sources being the throat-localised flux $\sim g_sN$ and the compensating negative tadpole elsewhere in the compact space. The 6d Greens function behaviour $\sim 1/r^4$ then implies that the typical variations of $(e^{-4A}-1)$ and hence of $(e^{2A}-1)$ on the scale of the Calabi-Yau are $\sim g_sN/r^4\sim N/\mathcal{V}^{2/3}$.

We deviate from \cite{Junghans:2022exo} in that we expect a scaling of the warping correction term with the Euler number $\chi(X)$. The reason is that we assume a slowly varying warp factor \cite{Gao:2020xqh} and hence do not expect a parametrically strong cancellation between the contributions from different regions. 

The correction to the Einstein-Hilbert term then translates into a Kahler potential correction and, analogously to the $\alpha'$ effect in the last term of \eqref{pureLVSpotential}, to a correction to the scalar potential:\footnote{Another important correction which is of the same form as BBHL was found in \cite{Minasian:2015bxa}. This can be taken into account in our analysis by replacing $\xi\to\xi+\xi'$ where $\xi'$ is computed in \cite{Minasian:2015bxa}.} 
\begin{equation}
    \delta V = \frac{15\,\xi\, N \,|W_0|^2}{8\sqrt{g_s}\,\mathcal{V}^{11/3}}\,\mathcal{O}(1) \,.
    \label{deltav}
\end{equation}
This correction in turn corrects the value of the (A)dS minimum and the stability condition of the dS minimum \cite{Junghans:2022exo}. We focus on the effect on the value of the potential at the minimum since the stability of the minimum depends crucially on the unknown sign of the correction \eqref{deltav}. 
A measure for parametric control is given by comparing the size of the correction to the scalar potential \eqref{deltav} and its value at the minimum \eqref{vads}.
Hence, we are prompted to consider the following improved version of \eqref{cN}:
\begin{equation}
    1= c_N \frac{10 \,a_s\,\xi^{2/3}}{(2\kappa_s)^{2/3}g_s} \frac{N}{\mathcal{V}^{2/3}} \,.
    \label{lowerbound}
\end{equation}
A large control parameter $c_N\gg 1$ again ensures a parametric suppression of the correction compared to the leading term\footnote{Note that, as described in \cite{Junghans:2022exo}, the dS minimum disappears if the correction \eqref{deltav} has an unfavourable sign and is too large. This correspond in our analysis to a value of $c_N<11C^\text{\,flux}/(18\xi)$ (see equation (5.8) in \cite{Junghans:2022exo}). Evaluating this for the explicit model considered there, the dS minimum breaks down for $c_N\simeq1.35$ (see equation (6.23)).}.

Now, we require the volume in \eqref{lowerbound} to be equal to \eqref{tauV} and replace subsequently $g_s$ by $\tau_s$ using \eqref{vts}. In doing so, we leave the combination $g_sM^2$ untouched since it represents an important parameter characterising the metastability of the throat. This leads to an equation for $N$ which is of the form $w\text{e}^w=x$, where
\begin{equation}
    w=-\frac{16\pi}{21g_s M^2}N\,,\quad\quad x =- \frac{3^{3/5}\pi^{9/35}\,a_0^{2/7}}{14\,\,2^{2/15}\, 5^{3/7}} \frac{\kappa_s^{2/7}\,\xi^{2/7}}{a_s^{3/7}\, c_{W_0}^{2/7}\, c_N^{3/7}} \frac{1}{(g_sM^2)^{1/7}}
    \,.
    \label{defofx}
\end{equation}

An equation of the above type is solved by $w=\mathcal{W}_{-1}(x)$, where $\mathcal{W}_{-1}(x)$ is a branch of the Lambert $\mathcal{W}$ function. We find the exact result
\begin{equation}
    N  = - \frac{21g_s M^2}{16\pi} \mathcal{W}_{-1}(x) = - \frac{21g_s M^2}{16\pi} \biggl( \ln(-x) +\ln(-\ln(-x)) + \mathcal{O}(1)\biggr)\,,
    \label{exactresult}
\end{equation}
where we expanded $\mathcal{W}_{-1}(x)$ around $x=0$. 
In order for the D3-tadpole in our compactification to cancel, we must at least have sufficient negative tadpole $Q_3$ from O3/O7-planes and D7-branes to cancel the flux in the throat
\begin{equation}
\label{eq:basicfluxtadpolecancellation}
    -Q_3 > N\,.
\end{equation}
In addition to \eqref{eq:basicfluxtadpolecancellation}, there is another lower bound on $-Q_3$ given by \cite{Denef:2004ze}\footnote{We thank Erik Plauschinn and the referee for making us aware of this important point.}
\begin{equation}
    -Q_3 \ge 4\pi\frac{g_s W_0^2}{2}\,,
    \label{denefdouglas}
\end{equation}
where the $4\pi$ arises since our superpotential $W$ is normalized differently compared to \cite{Denef:2004ze}. Expressing $W_0$ by \eqref{cw0} and subsequently $\mathcal{V}$ by \eqref{lowerbound} we find
\begin{equation}
    -Q_3 \ge \frac{c_N}{c_{W_0}} \frac{15\pi\xi}{4} N \equiv c_Q N\,.
    \label{denefdouglas1}
\end{equation}
Our main result is then that by filling in \eqref{exactresult} for $N$ in \eqref{eq:basicfluxtadpolecancellation} and \eqref{denefdouglas1}, we obtain two lower bounds on the D3-tadpole. For any given model and two prescribed values $c_N$ and $c_{W_0}$, we should check that the strongest of these bounds is satisfied. 

This result allows for a more compact formulation if we do not fix $c_N$ and $c_{W_0}$ but merely restrict them such that some minimal quality of control is ensured. Specifically, we demand $c_N\geq c_{N,{\rm min}}$ and $c_{W_0}\geq c_{W_0,{\rm min}}$, assuming also that the two values $c_{N,{\rm min}}$ and $c_{W_0,{\rm min}}$ are comparable. This is illustrated in Fig.~\ref{fig:control}, where the grey area represents the forbidden regime. The blue line in the plot separates the regimes where one or the other of the two bounds is stronger. It is described by $c_{W_0}/c_N=15\pi\xi/4$ or, equivalently, $c_Q=1$. Since $15\pi \xi/4$ is in general significantly larger than unity, the blue line cuts the horizontal rather than the vertical part of the boundary between the grey and white area.

Over the white part of the plot in Fig.~\ref{fig:control}, we have a function $-Q_3^*(c_N,c_{W_0})$ given by the r.h.~side of \eqref{eq:basicfluxtadpolecancellation} or \eqref{denefdouglas1}, whichever is largest. For a conservative constraint, we want to know the minimal value of that function. To determine this value, we first consider the area to the right of the blue line, where $c_Q<1$. It is immediately clear from eqs.~\eqref{defofx}-\eqref{eq:basicfluxtadpolecancellation} that, as indicated by the arrows, our function $-Q_3^*(c_N,c_{W_0})$ falls as we move to smaller $c_{W_0}$ at fixed $c_N$. Hence the minimum is attained on the blue line or to its left. On the contrary, in the region left of the blue line, i.e.~for $c_Q>1$, the value of $-Q_3^*(c_N,c_{W_0})$ generically falls if one moves to larger $c_{W_0}$. This requires a moment of thought: One first convinces oneself that, in this regime and with fixed $c_N$, the variation of the minimal tadpole reads $\delta[-Q_3^*(c_N,c_{W_0})]\sim \delta [(a+\ln(c_{W_0}))/c_{W_0}]$, with $a>0$. This follows from eqs.~\eqref{defofx}, \eqref{exactresult} and \eqref{denefdouglas1}, using also the leading logarithmic approximation in \eqref{exactresult}. It is now easy to see that, for sufficiently large $c_{W_0}$, the function $-Q_3^*$ always falls if $c_{W_0}$ grows. The requirement that $c_{W_0}$ is sufficiently large is always satisfied since we are outside the grey band. Thus, the arrows left of the blue line must indeed point to the right, and the minimum is hence on top of the blue line. Finally, restricting attention to points on the blue line, the value of $-Q_3^*$ falls as one moves towards the origin. 

As a result, the minimum is at the point where the blue line crosses into the grey area. This is specified by $c_N=c_{N,{\rm min}}$ and
\begin{equation}
    c_{W_0}^\ast = c_{N,{\rm min}} \frac{15\pi\xi}{4}\,.
    \label{cWstar}
\end{equation}
For simplicity, we drop the index `min' on $c_N$ in what follows.
Then, inserting \eqref{cWstar} together with $x$ into \eqref{exactresult}, we obtain the leading order result:

\begin{figure}
    \centering
    \includegraphics[scale=0.9]{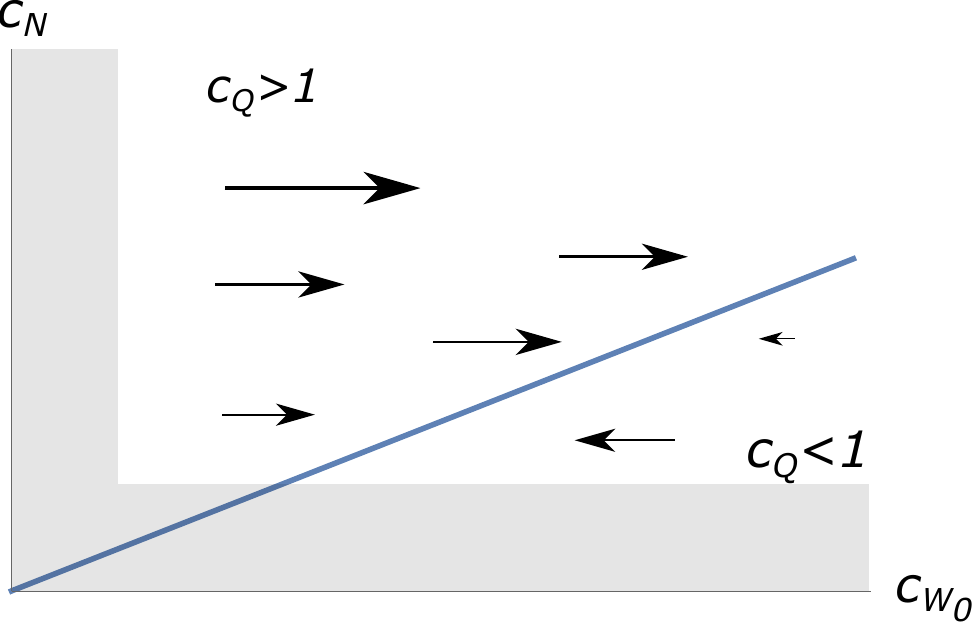}
    \caption{The parameter space of possible values of $c_N$ and $c_{W_0}$. In the grey area, one of these two control parameters is too small such that corrections are dangerously large. The blue line separates the regimes of validity of the two bounds in \eqref{eq:basicfluxtadpolecancellation} and \eqref{denefdouglas1}. The arrows specify how $c_{W_0}$ must change to decrease the lower bound $-Q_3^*(c_N,c_{W_0})$ at fixed $c_N$.}
    \label{fig:control}
\end{figure}

\begin{tcolorbox}
\textbf{The LVS parametric tadpole constraint:}\\
The D3 tadpole contribution $Q_3$ of O3/O7-planes and D7-branes must fulfil
\begin{equation}
    -Q_3 > N= N_\ast \biggl(\frac{1}{3}\ln N_\ast+ \frac{5}{3}\ln c_N + \ln a_s -\frac{2}{3}\ln \kappa_s +8.2  +\mathcal{O}(\ln(\ln))\biggr)\,,
    \label{eqtadpoleconstraint}
\end{equation}
where we defined $\quad N_\ast=9g_sM^2/(16\pi)$.
\end{tcolorbox}

Note that the subleading terms in \eqref{eqtadpoleconstraint} have a significant impact on the final bound one obtains. We only display the leading term to give the reader an intuition of how the bound scales with the parameters. However, to actually compute $N$ we always make use of the exact expression \eqref{exactresult}.

Thus, we require $Q_3$ to be negative and its absolute value to exceed the throat-flux $N$ since additional 3-form flux is needed to stabilize the complex-structure moduli. Let us discuss the implications of \eqref{eqtadpoleconstraint}. The constraint specifies the minimal tadpole needed in LVS constructions to avoid the singular-bulk problem, and to make sure that the corrections induced by a varying warp factor and higher F-terms are parametrically suppressed. The degree of parametric suppression can be chosen at will by specifying $c_N$. The parameter $c_N$ determines to which extent warping effects are suppressed and fixes, through \eqref{cWstar}, the optimal value of $W_0$.
The approximation of $x$ being close to zero improves with increasing $g_sM^2$ and increasing control parameters. Interestingly, the main expansion parameter $N_\ast$ coincides with the size of the small cycle $\tau_s$. 

An optimistic bound on $N$ is obtained by using the metastability constraint $g_sM^2\gtrsim 12$ by KPV \cite{Kachru:2002gs} (to be precise, $M>12$ combined with $g_s M\gtrsim 1$). Setting $\kappa_s=1$ 
\footnote{$\kappa_s$ is usually smaller than one, strengthening the bound. For example in the model discussed in \cite{Crino:2020qwk} $\kappa_s\approx 0.16$.} 
and $a_s=\pi/3$ for the moment, as well as using \eqref{exactresult} together with \eqref{cWstar}, we obtain for the minimal tadpole
\begin{equation}
    N\simeq 33.6 \qquad\text{for}\quad c_N=5\,\qquad\text{and}\,\qquad N\simeq 45.8 \qquad \text{for}\quad c_N=100\,.
\end{equation}
This is, however, probably too weak since it ignores backreaction. Indeed, the constraint comes from the process in which the $\overline{D3}$ polarises into a fluxed NS5, which slips over the equator of the $S^3$ at the tip of the throat. One expects that, as part of this decay process, the volume of the $S^3$ is driven to a smaller value, lowering the potential barrier for the decay.\footnote{
Note 
that this volume is {\it not} the same as the conifold modulus $Z$. Indeed, due to a cancellation between the $Z$-dependence of the warping and the $Z$-dependence of the internal metric at the tip of the warped throat, the physical volume of the $S^3$ at the tip is $Z$-independent. Correspondingly, the analysis of \cite{Scalisi:2020jal} does not find such a lowering of the potential barrier.}

An improvement has been suggested in \cite{Bena:2018fqc,Blumenhagen:2019qcg} (see also \cite{Douglas:2007tu,Douglas:2008jx}), where the conifold modulus is treated as a dynamical variable. In \cite{Scalisi:2020jal} this analysis was combined  with the polarization of the $\overline{D3}$ into an NS5-brane. As already discussed in \cite{Gao:2020xqh, Junghans:2022exo}, the reliability of these analyses is limited by the assumption that the warp factor behaves as $\exp(4A(0))\sim |Z|^{4/3}$, also off-shell. This is not obvious because $\exp(4A(0))\sim \exp(-8\pi K/3g_s M)$ is a priori a fixed number, determined by the flux choice. Put differently, it is not clear why the warp factor should go to zero as the $S^3$ at the tip of the throat is dynamically driven to zero radius. After all, the fluxes $K$ and $M$ determining $\exp(4A(0))$ \`{a} la Klebanov-Strassler are discrete.

Nevertheless, we may take for the moment the value $g_sM^2\approx 46$ obtained in \cite{Bena:2018fqc,Blumenhagen:2019qcg} as a (probably rather conservative) lower bound and insert it together with \eqref{cWstar} in \eqref{exactresult}. Choosing again $\kappa_s=1$ and $a_s=\pi/3$, we find
\begin{equation}
    N\simeq 132.9\qquad\text{for}\quad c_N=5\qquad\text{and}\qquad N\simeq 179.9 \qquad \text{for}\quad c_N=100\,.
\end{equation}
Combined with the fact that, in the explicit LVS models considered to date the maximal tadpole is $\sim 150$ \cite{Crino:2020qwk}, we see that the tadpole constraint is neither harmless nor deadly.

The natural way to ensure the constraint \eqref{eqtadpoleconstraint} is satisfied is to search for LVS compactifications with a large D3-tadpole. We will discuss the prospects for finding such examples and stabilizing all moduli in Sect.~\ref{sec:tadpoleconjecture}. In addition, one may try to study the LVS in more detail in the regime where the control parameters $c_N$ and $c_{W_0}$ are only moderately large. To convince oneself that the construction remains reliable, one would need to make sure that even higher order corrections of the same type do not blow up. Interestingly, \eqref{eqtadpoleconstraint} increases for small $\kappa_s$ and is independent of the Euler number. This appears to partly differ from the trend observed in example studies in \cite{Crino:2020qwk}.

\section{Discussion of the LVS Tadpole Constraint}
\label{sec:tadpoleconjecture}

\subsection{Interplay with the Tadpole Problem}

Our central result \eqref{eqtadpoleconstraint} provides a quantitative lower bound on the flux-induced D3 tadpole. The tadpole conjecture \cite{Bena:2020xrh,Bena:2021wyr,Plauschinn:2021hkp,Lust:2021xds,Bena:2021qty} on the other hand states that, for a large number $h^{2,1}$ of complex structure (CS) moduli, the tadpole of the 3-form flux required to stabilize all of them in a non-singular regime grows faster than the negative tadpole of the corresponding orientifold geometry. If the conjecture were true, it would be logical to focus on LVS models with a large tadpole and a small number of moduli, escaping the allegedly dangerous asymptotic regime. But this may not be possible since, according to known examples, a large negative tadpole comes with a complicated topology (as discussed in more detail in Sect.~5 of \cite{Gao:2020xqh}) and hence many moduli. Indeed, one would expect that a single involution of a certain geometry has many O3 fixed points precisely because the topology is complicated. Similarly, while even a single O7 plane with its D7 branes can produce a large tadpole, this tadpole is tied to its topology. It grows if the 7-brane topology becomes complicated, which is in general associated with the presence of many brane-deformation moduli. These, in turn, require more flux to be stabilized.

It then seems that the allowed LVS vacua are constrained to an interval of permitted $Q_{3}$, with \eqref{eqtadpoleconstraint} providing a lower bound and the parametrics of the tadpole conjecture providing an upper bound above which it becomes hard to stabilize all the CS moduli. One might worry that further analyses will demonstrate that this interval shrinks to zero size, casting all LVS de Sitter vacua into the swampland. Indeed, this concern was already raised in \cite{Bena:2020xrh}, though without deriving a quantitative LVS tadpole constraint. However, things need not be all doom and gloom:

We recall an important but perhaps underappreciated condition for the tadpole conjecture to apply, namely that the stabilized geometry is required to be smooth. For example, in the simple $K3\times K3$ example it is possible to stabilize all CS moduli (including D7 brane moduli in the language of F-theory) if singular geometries are permitted \cite{Dasgupta:1999ss, Aspinwall:2005ad, Braun:2008pz, Braun:2014ola}. According to the detailed study of \cite{Bena:2020xrh}, developing the method of \cite{Braun:2008pz}, this fails if one attempts to avoid the gauge enhancement through singularities. Nevertheless, this failure allows for the following more optimistic view on the possible string landscape of de Sitter vacua: The tadpole conjecture says that all vacua in the landscape at large number of CS moduli have singular internal geometries. One may embrace this statement together with the fact that the singularities will in general produce enhanced nonabelian gauge symmetries. The dynamics of those gauge theories together with SUSY breaking might stabilize the new moduli associated to the singularity. If this happens only at low energies, the nonabelian gauge theories make our compactification more interesting phenomenologically. The tadpole conjecture then tells us that the de Sitter vacua that do exist in the landscape are on average much more interesting phenomenologically then one a priori expects.

\subsection{Overcoming the LVS Tadpole Constraint and further challenges}

With these considerations in mind, let us look at which D3-tadpole has been achieved in IIB string theory and what the prospects for realizing higher tadpoles might be.

In IIB string theory, the $D3$ tadpole is generally given by
\begin{equation}
\label{eq:D3tadpole}
 N_{D3}+\frac{N_{\text{flux}}}{2}+N_{\text{gauge}}=\frac{N_{O3}}{4}+\frac{\chi(D_{O7})}{12}+\sum_a N_a\, \frac{\chi(D_{a})+\chi(D'_{a})}{48} \equiv -Q_3\,,
\end{equation}
with $N_{\text{flux}}=\frac{1}{(2\pi)^4 \alpha^{'2}}\int H_3\wedge F_3 $, $N_{\text{gauge}}=-\sum_{a} \frac{1}{8\pi^2} \int_{D_a} \text{tr}{\cal F}_a^2$, where ${\cal F}_a$ is the gauge flux turned on on each of the divisors $D_a$, and $N_a$ the number of D7 branes (their orientifold images) wrapping the divisor $D_a$ ($D'_a$). Note that a throat with fluxes $K$ and $M$ contributes an amount of $2N=2KM$ to the quantity $N_{\rm flux}$ in \eqref{eq:D3tadpole}. The reason is that a throat in the orientifold corresponds to two throats in the Calabi-Yau. If the D7 brane tadpole is cancelled locally by placing eight $D7$ branes on top of the $O7$ plane, \eqref{eq:D3tadpole} simplifies to \be
\label{d3tadsimple}
      N_{D3} + \frac{N_{\text{flux}}}{2}+ N_{\rm gauge}= \frac{N_{O3}}{4}+\frac{\chi(D_{O7})}{4}\, .
\ee

Most concrete type-IIB orientifolds considered in the literature are based on Calabi-Yaus with a small number of Kahler moduli $h^{1,1}(X)$. The highest negative tadpole values of explicitly considered models we are aware of are $Q_3 = -149$ \footnote{This is the model considered in \cite{Junghans:2022exo}.} \cite{Crino:2020qwk} and $-104$ \cite{Louis:2012nb}, both in models with $h^{1,1}(X) = 2$ and based on reflection involutions. More generally in type-IIB orientifolds with locally cancelled D7 tadpole, the negative contribution to the D3 tadpole can be bounded from above by the Lefschetz fixed point theorem \cite{Collinucci:2008pf, Carta:2020ohw, Bena:2020xrh}:\footnote{We thank Jakob Moritz for pointing this out to us.}
\begin{equation}
    -Q_3\le 1 + \frac{1}{2} \left( h^{1,1} + h^{2,1} \right) \,.
\end{equation}
Using the largest Hodge numbers from the Kreuzer-Skarke dataset \cite{Kreuzer:2000xy}, this gives, $-Q_3\le 252$.

\begin{itemize}
\item {\it Exploiting non-local D7-tadpole cancellation}: In the models of \cite{Louis:2012nb,Crino:2020qwk}, the D7 tadpole is cancelled locally, i.e. eight D7 branes sit on top of each O7. This restricts the possibilities of getting a large tadpole: Indeed, according to \cite{Carta:2020ohw}, canceling the D7 tadpole locally in smooth and favorable Complete Intersection Calabi-Yau three-folds (CICYs) with both reflections and simple divisor exchange involutions gives even tadpole values in the range $ [-36, 4] $. By contrast, non-local D7-tadpole cancellation produced values in the larger range $[-132, -12 ]$. Moreover, the study of toric orientifold Calabi-Yaus with $h^{1,1} \le 6$ in \cite{Altman:2021pyc} gave $Q_3 \in [ -30,0]$. This was based only on the divisor exchange involution and local D7-tadpole cancellation. Models with reflection involution in toric Calabi-Yaus are still under study but we expect that in this class the tadpole will increase significantly. The reason is that, in the examples of \cite{Louis:2012nb, Crino:2020qwk} with reflection involution the tadpole is already as large as $104$ and $149$ even for $h^{1,1}(X) = 2$. Thus, one may hope that extending this to the range $h^{1,1}(X)\le 6$ much larger tadpoles can arise.

\item {\it Using divisors $D_a$ with large Euler number}\,:  (a) From \eqref{eq:D3tadpole} one can see another reason for the limited tadpole range arising in the study of \cite{Carta:2020ohw}. It is the fact that in CICYs the topology of each divisor is relatively simple and the largest Euler number of a divisor is modest:  $\chi(D_a)_{max} = 80 $ \cite{Carta:2020ohw, Carta:2022oex}. By contrast, in the toric setting even with $h^{1,1}(X) \le 6$, the highest Euler number that gives an integer contribution to the tadpole is $\chi(D_a)_{max} = 504 $ \cite{Altman:2021pyc}. Even if we only consider a reflection on this divisor and cancel the D7-tadpole locally, it will contribute $-126$ to the tadpole. A distribution of Euler numbers of divisors in the orientifold Calabi-Yau database of \cite{Altman:2021pyc} is displayed in Fig.~\ref{fig:dist}.
\begin{figure}[ht]
\centering
\includegraphics[width=10cm]{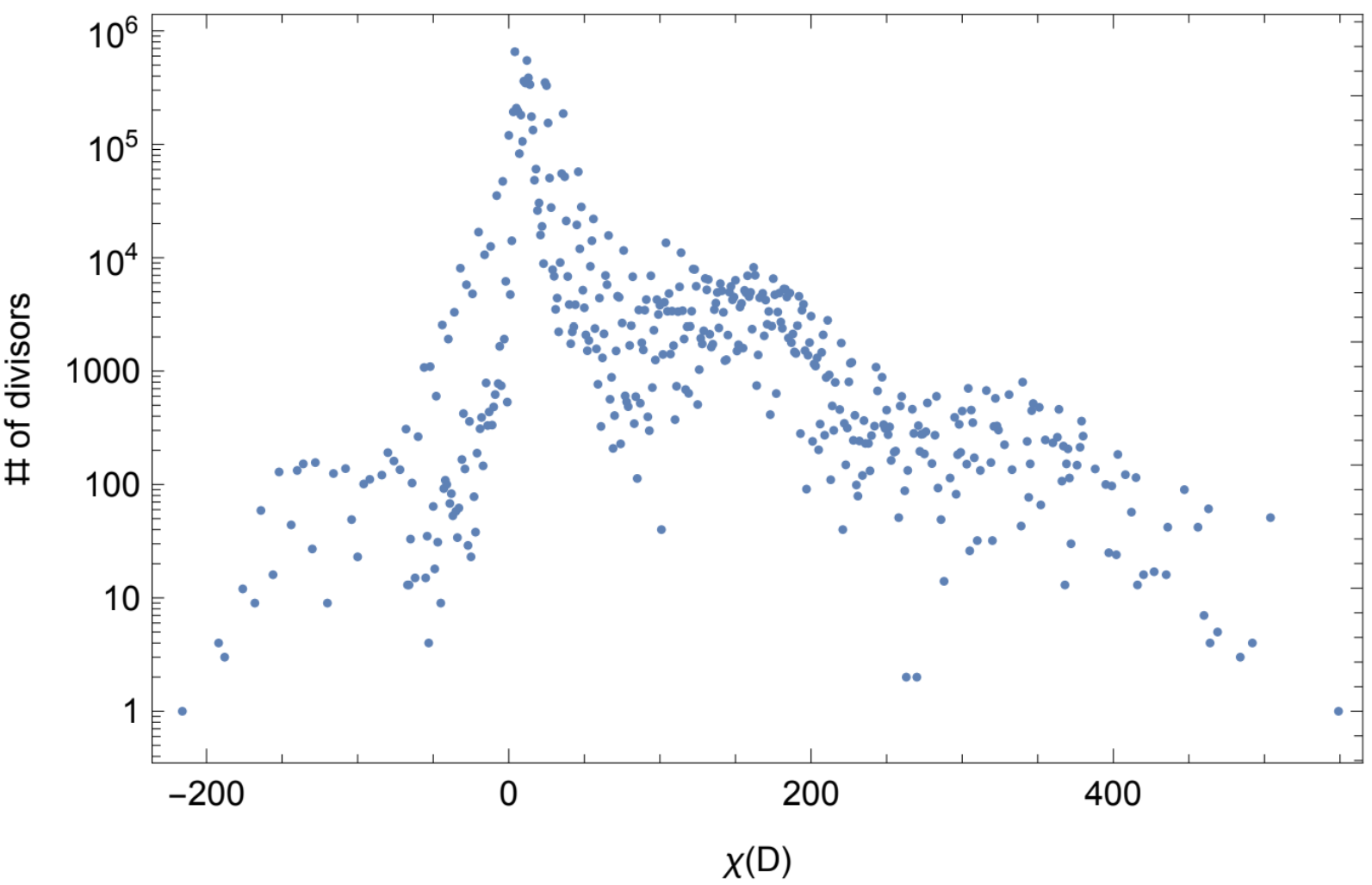}
\caption{Log distribution of the Euler number of individual divisors $\chi(D)$ in toric orientifold Calabi-Yaus with $h^{1,1}(X) \leq 6$.}
\label{fig:dist}
\end{figure}
$\,$(b) Another way to enlarge the Euler number of $D_a$ is by considering a D7-brane wrapping a combination of divisors. This relies on detailed model building and represents an interesting project for the future.

\item {\it Large  $h^{1,1}(X)$}:  Increasing $h^{1,1}(X)$ enriches the possibilities for D7-brane model building and hence should make it possible to find larger Euler number for $D_a$ and hence larger $N_{O3}$. For CICYs, the largest $h^{1,1}$ is just $19$ \cite{Candelas:1987kf} while for toric Calabi-Yau this number increases to $h^{1,1}(X)=491$ \cite{Kreuzer:2000xy}. Although so far no database for orientifolds of toric Calabi-Yaus with higher $h^{1,1}(X)$ exists, progress based on large cluster computation and machine learning has been made \cite{Gao:2021xbs}. We note that in the search for large gauge groups, a correlation between the rank and $h^{1,1}$ has been observed \cite{Louis:2012nb}. The concrete model studied in that paper has only $h^{1,1}=2$ and a maximal negative tadpole of $-92$, but the expectation is that models with much larger $h^{1,1}$ and D7 brane numbers going into the hundreds exist. While this raises the hope that very large negative tadpoles can also be achieved by the detailed study of the corresponding geometries, one in general pays the price of large $h^{1,1}$. This means many Kahler moduli, the stabilization of which may in itself be non-trivial in the LVS context, as we discuss below.

\item {\it Increasing $N_{O3}$}: Finally, an obvious way forward is to search for models with more O3 planes. In the particularly tractable setting of CICYs, see e.g.~\cite{Candelas:1987kf, Anderson:2017aux}, the multiplicities of O3 and O7 planes have recently been studied in \cite{Carta:2020ohw}. Tadpoles in the range $[ -132,-12 ]$ were found. An interesting aspect uncovered in this context is that the resolution branch of each conifold singularity on an O7 plane produces one extra O3. Although in \cite{Carta:2020ohw} this did not lead to higher tadpoles, this might change in more general settings, such as toric Calabi-Yaus. Moreover, the existence of this method for increasing the number of O3 planes, possibly in the range of $N_{O3} \sim 100$, is interesting in itself.
\end{itemize}

If it should nevertheless prove impossible to construct perturbative type-IIB models with a sufficiently high tadpole to be in control according to \eqref{eqtadpoleconstraint} and with all CS-moduli stabilized in an acceptable manner, another option is to consider proper F-theory compactifications\footnote{The F-theory analogue of the crucial correction \eqref{R4} was derived in \cite{Minasian:2015bxa}.} \cite{Vafa:1996xn, Weigand:2010wm}. The immediate advantage is that Calabi-Yau 4-folds with very large Euler number $\chi_4$ are known, leading to negative D3 tadpoles as high as $N=\chi_4/24=75852$ \cite{Klemm:1996ts, Candelas:1997eh}. This appears to be sufficient to easily satisfy the constraint  \eqref{eqtadpoleconstraint} with excellent control. But the extension of the discussion to F-theory models has its price:

On the one hand, the key parametric relations $\ln{\cal V}\sim \tau_s\sim 1/g_s$ demand that we realise a small value of $g_s$ to achieve a large volume. In F-theory, $g_s$ is replaced by an appropriate average of the exponential of the dilaton, $g_s \equiv \langle e^\phi \rangle$. Studying whether a small $g_s$ can in this way be implemented in large-tadpole F-theory geometries is an interesting task. 

On the other hand, a 4-fold with large Euler number implies, in the smooth case, that the F-theory model has either many Kahler moduli or many complex structure moduli or many 7-brane deformations (cf.~Sect.~5 of~\cite{Gao:2020xqh}, based on~\cite{Denef:2008wq, Collinucci:2008pf, Braun:2008ua}). The latter two options potentially run into trouble with the tadpole conjecture (though this may be acceptable if one allows for singular geometries). The former option, where the number of Kahler moduli is large, potentially has a problem with the LVS moduli stabilization procedure. Indeed, the standard approach to stabilizing non-blowup moduli (other than the volume) in the LVS is through loop corrections \cite{Cicoli:2007xp,Cicoli:2008va}. The dependence of such corrections on ratios of 4-cycle volumes as well as the complex-structure moduli dependence of the prefactors is not known except for simple torus orbifold geometries \cite{Berg:2005ja, Berg:2007wt} (see \cite{Gao:2022uop} for an attempt to make progress). This may not be a problem as a matter of principle since, generically, one expects that all Kahler moduli will somehow be stabilized by loop effects. But clearly it {\it is} a problem if one wants to establish rigorously the existence of at least one explicit LVS de Sitter model. Hence, the way forward may be not to rely on loops but rather demand that all ratios of non-blowup cycles are stabilized by $D$-term constraints 
\cite{Jockers:2005zy, Haack:2006cy, Blumenhagen:2007sm}. In this case, achieving a purely algebraic and fully explicit construction in the near future appears more plausible.

The correction we have focused on in this note is not the only correction relevant for control in the LVS. As emphasized in \cite{Junghans:2022exo}, further potentially dangerous corrections are associated with logarithmic terms. There are two types:

One of them \cite{Conlon:2010ji} derives from the field redefinition $\tau'_s \equiv \tau_s + \alpha \log (\mathcal{V})$, where $\tau_s'$ rather than $\tau_s$ is now the real part of the proper Kahler variable. The Kahler potential then reads 
\begin{equation}
    K \sim - 2\log\left[ \tau_b^{3/2}-\left(\tau_s'-\alpha \log \left(\mathcal{V}\right)\right)^{3/2} + .. \right] + .. \,,
\end{equation}
with the physical small-cycle volume being $(\tau_s'-\alpha \log (\mathcal{V}))$. It is at the moment not entirely settled if and when such a correction appears in string compactifications. A surprising aspect is the implicit claim that, at ${\cal V}\to \infty$, the correction grows in importance. For example, the size of non-perturbative effects $\sim \exp(-a_s\tau_s')$ now has a very different relation to the physical volumes of small cycle and Calabi-Yau.

Another type of correction, corresponding to a replacement $\mathcal{V}\to \mathcal{V}+\alpha\log(\mathcal{V})$, has been proposed to appear through the interplay of the $R^4_{10}$ and D7/O7 branes \cite{Antoniadis:2019rkh}. An interesting aspect of this proposal is that, while it indeed threatens parametric control of the LVS, it could instead allow for alternative ways to realize a de Sitter minimum.

The correction \eqref{deltav} we have focused on in this note and log corrections are not the only potentially relevant corrections for the LVS. As recently highlighted in \cite{Junghans:2022exo}, a host of further corrections are interesting or even potentially dangerous. This requires more research. In this context, in \cite{Gao:2022uop}, we give a general discussion of loop corrections to the Kahler moduli Kahler potential. While important in many ways, we do not see that these corrections endanger the basic setting of the LVS.

Finally, we emphasize again that the tadpole constraint \eqref{eqtadpoleconstraint} is tied to the $\overline{\rm D3}$-uplift. Hence, even if this constraint turned out to remain problematic for the LVS, other uplifting mechanisms could work more successfully, especially if they do not require a very large tadpole. One example is the $D$-term uplift, where the positive energy arises because a D7-brane matter-field is unable to satisfy simultaneously a $D$-term and an $F$-term constraint \cite{Achucarro:2006zf,Cremades:2007ig}. Implementing this consistently in the stringy geometry leads to the concept of a T-brane uplift \cite{Cicoli:2015ylx}. 
Another suggestion is the Winding uplift \cite{Hebecker:2020ejb, Carta:2021sms} (with the idea going back to \cite{Hebecker:2015rya}), which uses the possibility to create a perturbatively flat direction in the complex-structure moduli space by an appropriate flux choice \cite{Hebecker:2015rya}. The non-perturbatively small potential along this flat valley can then arguably be tuned such that a local metastable minimum with non-zero $F$-term arises. In this setting, the tuning power of the complex-structure landscape could help to realize an exponentially small uplift without a warped throat.

\section{Summary}
The core result of this short note is \eqref{eqtadpoleconstraint}, which bounds the D3 tadpole required in LVS models from below. The bound comes from demanding that warping corrections in the bulk, associated with the Klebanov-Strassler throat housing the anti-D3 brane uplift, are under control. While this constrains the set of geometries suitable for an LVS with a controlled uplift, we so far see no catastrophe that places all LVS vacua in the swampland. Rather, the bound poses a challenge for LVS model-building which one can attempt to overcome by searching for compactifications with a sufficiently large D3-tadpole. While the tadpole conjecture poses an additional challenge in this context, we do not currently see this as impossible to overcome. In type-IIB a tadpole of 149 has been achieved \cite{Crino:2020qwk}, but, as we argued above, much larger tadpoles may be possible. In F-theory, the largest available tadpole of $75852$ \cite{Klemm:1996ts, Candelas:1997eh} easily satisfies our bound \eqref{eqtadpoleconstraint} with excellent control. However, then one faces other challenges: One either has to realize a small averaged $g_s$ in F-theory or, alternatively, one has to replace the control parameter $1/g_s$ by a topology which makes the coefficient of the $\alpha'^3$ correction appropriately large. Moreover, stabilizing all moduli in a controlled manner in geometrically complicated models with high tadpole may introduce new difficulties. Finally, a key quantity determining the bound of the D3 tadpole \eqref{eqtadpoleconstraint} is $g_sM^2$, the minimal value of which is still under debate. In order to see how constraining the bound for the LVS really is, it would be essential to know this value as precisely as possible.

We hope that this note has highlighted a number of directions for future work which can help to clarify whether LVS de Sitter vacua exist.

\section*{Acknowledgements}

We thank Daniel Junghans, Jakob Moritz, Erik Plauschinn, Andreas Schachner, Pramod Shukla and Pablo Soler for valuable comments on an earlier version of this note.

X.G. was supported in part by the Humboldt Research Fellowship and NSFC under grant numbers 12005150. This work was supported by the Graduiertenkolleg ‘Particle physics beyond the Standard Model’ (GRK 1940) and the Deutsche
Forschungsgemeinschaft (DFG, German Research Foundation) under Germany’s Excellence Strategy EXC 2181/1 - 390900948 (the Heidelberg STRUCTURES Excellence Cluster).

\appendix

\section{Derivation of the \boldmath{$\overline{D3}$} Uplift Potential}\label{upliftpotential}

In this appendix we derive the uplift potential directly from the Klebanov-Strassler (KS) geometry \cite{Klebanov:2000hb}, using also \cite{Klebanov:2000nc,Herzog:2001xk}. This provides a more precise derivation of the (non-backreacted) uplift potential  
than we are aware of elsewhere in the literature. 

The warp factor in string frame far away from the tip of the throat is given by the following two equivalent expressions. Firstly, it can be parametrized by the flux $N=KM$ and the radial distance from the tip $r=r_\text{max}$ at which the total flux in the throat equals $N$. At $r_\text{max}$, the throat is glued into the compact Calabi-Yau. This yields \cite{Klebanov:2000nc,Herzog:2001xk}
\begin{equation}
    h_N (r) = \frac{27\pi \alpha'^2}{4r^4} \left( g_s N + \frac{3(g_s M)^2}{2\pi}\ln(r/r_\text{max}) + \frac{3}{8\pi}(g_sM)^2 \right)\,.
    \label{hN}
\end{equation}
Secondly, we may use the KS warp factor $h_\varepsilon^\text{KS} (r)$ which describes both the UV and the smooth, IR region and which is parametrized by the conifold resolution parameter $\varepsilon$ \cite{Klebanov:2000hb, Herzog:2001xk}. In the UV, it has the approximate form
\begin{equation}
    h_\varepsilon^\text{KS} (r) \simeq \frac{81 (\alpha'g_s M )^2}{8r^4}\ln\left( \left( \frac{2^{5/3}}{3} \right)^{1/2} \frac{r}{\varepsilon^{2/3}} \right)\,.
    \label{hepsilon}
\end{equation}
Comparing \eqref{hN} and \eqref{hepsilon} determines $r_\text{max}$ as a function of $\varepsilon$ to leading order to be 
\begin{equation}
    r_\text{max} =\left( \frac{3}{2^{5/3}} \right)^{1/2} \varepsilon^{2/3}\, \text{e}^ {\frac{2\pi K}{3 g_s M}}\,.
    \label{rmax}
\end{equation}
Moreover, from KS, the warp factor at the tip is known. In string frame, it reads
\begin{equation}
    h_\varepsilon^\text{KS} (0) = \frac{2^{2/3} a_0 (g_sM\alpha')^2}{\varepsilon^{8/3}}\,,
\end{equation}
with the constant $a_0\approx 0.71805$.

With this in hand, we can calculate the ratio of the warp factors at the tip and far away from the tip at $r=r_\text{max}$. This yields the relative warping needed for the uplift potential:
\begin{equation}
    \frac{ h_\varepsilon^\text{KS} (0)}{h_\varepsilon^\text{KS} (r=r_\text{max})} = \frac{a_0}{3\,2^{2/3}\pi} \, \frac{g_s M^2}{N }\text{e}^ {\frac{8\pi K}{3 g_s M}}\,.
    \label{ratioofh}
\end{equation}
So far, we neglected the additive, constant contribution $\sim \mathcal{V}^{2/3}$ to the warp factor \cite{Giddings:2005ff}\footnote{The warp factor is determined by a Poisson equation and can hence be always shifted by a constant.} which becomes dominant at sufficiently large volume. Specifically, we replace the KS warp factor according to
\be
h_\epsilon^{\rm KS}(r) \qquad \to\qquad ({\cal V}_s/{\cal V}_{s,0})^{2/3} \,h_\epsilon^{\rm KS}(r_{\rm max}) \,+\, h_\epsilon^{\rm KS}(r)\,. 
\ee
Here we normalized the constant term by introducing a string-frame, fiducial volume ${\cal V}_{s,0}$. Our definition implies that, if ${\cal V}_s={\cal V}_{s,0}$, the strong-warping region is set by $r \lesssim r_{\rm max}$. This is the situation when a strongly warped throat is glued into a weakly warped, compact CY. For larger ${\cal V}_s$, a weakly-warped conical region develops above the throat (see e.g.~\cite{Brummer:2005sh}). From all of this, it should be clear that ${\cal V}_{s,0}$ must be chosen such that typical length scales in the CY match typical length scales of the $T^{1,1}$ at $r_{\rm max}$. Hence, for ${\cal V}_s\gg {\cal V}_{s,0}$, the full, inverse warp factor at the tip is given by
\begin{equation}
    \text{e}^{4A(0)} = \left(\frac{\mathcal{V}_s}{\mathcal{V}_{s,0}}\right)^{2/3} \frac{h_\varepsilon^\text{KS} (r=r_\text{max})}{h_\varepsilon^\text{KS} (0)}\, = \left(\frac{\mathcal{V}_s}{( \left.\text{Vol}(T^{1,1})\right|_{r_\text{max}})^{6/5}}\right)^{2/3} \frac{h_\varepsilon^\text{KS} (r=r_\text{max})}{h_\varepsilon^\text{KS} (0)}\,.
    \label{warpfactor}
\end{equation}
The volume scaling of this expression matches the corrected volume scaling of \cite{Kachru:2003sx}. 

To obtain the final formula of the uplift potential, we hence need to determine the volume of the $T^{1,1}$ at $r_\text{max}$. For this we consider the string frame metric of the conifold far away from the tip
\begin{equation}
    \dd s_{10}^2 = h_N^{-1/2}(r) \, \eta_{\mu\nu}\dd x^\mu \dd x^\nu + h_N^{1/2}(r)\left( \dd r^2 + r^2 \dd s_{T^{1,1}}^2 \right)\,.
\end{equation}
Using \eqref{hN} and Vol$\,T^{1,1}=16\pi^3/27$ from \cite{Gubser:1998vd}, we find
\begin{equation}
    \mathcal{V}_{s,0}^{2/3} = \frac{3^{3/5}}{2^{14/5}\,\pi^{3/5}}\left( g_s N +\frac{3}{8\pi} (g_sM)^2 \right)\,.
    \label{v23}
\end{equation}
This can be plugged into \eqref{warpfactor} which then yields the uplift potential
\begin{equation}
    V_{\text{uplift}} = V_{\overline{D3}} = 2 T_{D3} e^{4 A(0)} = \frac{\left( 3^2\,\pi^3\, 2^{22/3} \right)^{1/5}}{a_0} \,\frac{\text{e}^{-\frac{8\pi K}{3g_s M}}}{g_s M^2\mathcal{V}^{4/3}}
    \,.
\end{equation}
Here we only used the leading, first term on the r.h.~side of \eqref{v23}. Note that the $g_sM^2$ dependence of our the uplift potential agrees with the results of the supergravity-based approach of \cite{Douglas:2007tu,Douglas:2008jx,Bena:2018fqc}. At the moment, we do not understand why this agreement occurs in spite of the absence of a fiducial volume ${\cal V}_{s,0}$ in these references. In our approach, ${\cal V}_{s,0}$ contributes essentially to the final result.

\bibliographystyle{JHEP}
\bibliography{refs}

\end{document}